\documentclass[12pt,a4,english]{article}

\usepackage{bm}
\usepackage{fontspec}
\usepackage{units}
\usepackage{mathtools}
\usepackage{amsmath}
\usepackage{amssymb}
\usepackage{mathtext}
\usepackage{graphicx}
\usepackage{esint}
\usepackage[english]{babel}
\usepackage[
            backend=biber,
            style=gost-numeric, 
            movenames=false,    
            language=auto,
            autolang=other,
            otherlangs=true,
            blockpunct=space,
            citeurl=false,
            citedoi=false,
            biburl=false,
            bibdoi=true,
            bibisbn=false,
            bibeprint=false,
            sorting=none
            ]
            {biblatex}

\usepackage[autostyle]{csquotes}
\usepackage{amsmath,amsfonts,amssymb}
\usepackage[mathscr]{eucal}
\usepackage{subcaption}
\usepackage{caption}
\usepackage{float}

\usepackage[%
    colorlinks=true,
    linkcolor=black,
    citecolor=black,
    urlcolor=blue,
    filecolor=black,
    pdftitle={Multifluid Hydrodynamic Simulation of Metallic-Plate Collision Using the VOF Method},
    pdfauthor={Fedor Belolutskiy},
    pdfkeywords={compressible multiphase flows, multifluid simulation, VOF method, volume of fluid, collision of metal plates, equations of state, three-phase model, condensed phases, shock waves, unloading waves},
    pdfproducer={LaTeX},
    pdfcreator={xelatex}
]{hyperref}

\DeclareDelimFormat{multicitedelim}{\addcomma\space}

\AtEveryBibitem{%
  \clearfield{month}
  \clearfield{day}
  \clearfield{endmonth}
  \clearfield{endday}
}

\usepackage{marginnote} 
\reversemarginpar

\let\starcaption\caption  

\makeatletter
\renewcommand{\fnum@figure}{\textbf{Fig.\ \thefigure}}
\makeatother

\usepackage{titlesec}
\renewcommand{\thesection}{\arabic{section}}
\titleformat{\section}
  {\normalfont\centering} 
  {\thesection.}                         
  {0.33em}
  {\MakeUppercase}                                    
\titlespacing*{\section}
  {0pt}                                 
  {4mm}                                 
  {2mm}

\titleformat{name=\section,numberless}  
  {\normalfont\centering}
  {}                                    
  {0pt}                                 
  {\MakeUppercase}
\titlespacing*{\section}
  {0pt}{4mm}{2mm}
\usepackage{indentfirst} 

\renewcommand{\thesubsection}{\thesection.\arabic{subsection}}
\titleformat{\subsection}
  {\normalfont\centering}
  {\textit{\thesubsection.}}
  {0.33em}
  {\textit}
\titlespacing*{\subsection}
  {0pt}
  {1\baselineskip}
  {0pt}

\usepackage{libertinus}
\usepackage{setspace}
\usepackage[top=2.9cm, bottom=2cm, left=2.4cm, right=2.4cm]{geometry}

\makeatletter
\renewcommand*{\@biblabel}[1]{\hfill#1.}
\makeatother

\newcommand{\mus}{\upmu s}

\AtBeginBibliography{\normalfont}
\DeclareNameAlias{author}{given-family} 
\DeclareNameAlias{editor}{given-family}
\DeclareNameAlias{translator}{given-family}
\DeclareNameAlias{heading}{given-family}

\DeclareFieldFormat[article,incollection,inproceedings,unpublished]{title}{``#1''}
\DeclareFieldFormat[thesis]{title}{``#1''}

\DeclareFieldFormat[article]{volume}{\textbf{#1}}
\DeclareFieldFormat[article]{number}{(#1)} 

\DeclareFieldFormat{pages}{#1}
\newbibmacro*{mypages}{%
  \printfield{pages}}

\newbibmacro*{issue+date}{%
  \printtext[parens]{\printdate}%
  \newunit}

\newbibmacro{doi}{%
  \printfield{doi}%
}

\DeclareFieldFormat{doi}{%
  \normalfont
  \ifhyperref
    {\\\href{https://doi.org/#1}{https://doi.org/#1}}%
    {\\https://doi.org/#1}%
  \adddot} 

\DeclareFieldFormat[article,incollection,inproceedings,unpublished]{title}{``#1\addcomma''}
\DeclareFieldFormat[inbook]{title}{``#1\addcomma''} 
\DeclareFieldFormat[thesis]{title}{``#1\addcomma''} 

\DeclareFieldFormat[book]{title}{\textit{#1}}
\DeclareFieldFormat[collection]{title}{\textit{#1}}
\DeclareFieldFormat[proceedings]{title}{\textit{#1}}
\DeclareFieldFormat[incollection]{booktitle}{\textit{#1}}
\DeclareFieldFormat[inproceedings]{booktitle}{\textit{#1}}
\DeclareFieldFormat[inbook]{booktitle}{\textit{#1}}

\DeclareBibliographyDriver{article}{%
  \usebibmacro{bibindex}%
  \usebibmacro{begentry}%
  \usebibmacro{author/translator+others}%
  \setunit{\addcomma\space}\newblock 
  \usebibmacro{title}%
  \newblock
  \usebibmacro{journal}%
  \setunit{\addspace}
  \printfield{volume}%
  \printfield{number}
  \setunit{\addcomma\space}
  \usebibmacro{mypages}
  \setunit{\addspace}%
  \usebibmacro{issue+date}%
  \setunit{\adddot\space}%
  \usebibmacro{doi}%
  \usebibmacro{finentry}}

\renewbibmacro{in:}{%
  \ifentrytype{article}{}{\printtext{\bibstring{in}\intitlepunct}}}

\DefineBibliographyStrings{english}{
  pages = {},
  page = {},
}
\DefineBibliographyStrings{russian}{
  pages = {},
  page = {},
}

\DeclareNameWrapperFormat{author}{\normalfont #1}
\DeclareNameWrapperFormat{editor}{\normalfont #1}

\DeclareDelimFormat[bib]{finalnamedelim}{%
  \ifnumgreater{\value{liststop}}{2}{\finalandcomma}{}%
  \addspace\bibstring{and}\space}

\renewbibmacro*{name:family-given}[4]{%
  \ifuseprefix
    {\usebibmacro{name:delim}{#3#1}%
     \usebibmacro{name:hook}{#3#1}%
     \ifblank{#3}{}{#3 }%
     \mkbibnamegiven{#2}\addspace
     \mkbibnamefamily{#1}%
     \ifblank{#4}{}{ #4}}
    {\usebibmacro{name:delim}{#1}%
     \usebibmacro{name:hook}{#1}%
     \mkbibnamegiven{#2}\addspace
     \mkbibnamefamily{#1}%
     \ifblank{#4}{}{ #4}}%
}

\DeclareBibliographyDriver{book}{%
  \usebibmacro{bibindex}%
  \usebibmacro{begentry}%
  \normalfont\printnames{author}%
  \setunit{\normalfont\addcomma\space}\newblock
  \usebibmacro{title}
  \setunit{\addspace}%
  \printtext{(}\printlist{publisher}%
  \setunit{\normalfont\addcomma\space}%
  \printlist{location}%
  \setunit{\normalfont\addcomma\space}%
  \printfield{year}%
  \printtext{)}%
  \usebibmacro{finentry}}

\DeclareBibliographyDriver{incollection}{%
  \usebibmacro{bibindex}%
  \usebibmacro{begentry}%
  \normalfont\printnames{author}%
  \setunit{\normalfont\addcomma\space}\newblock
  \usebibmacro{title}
  \newblock
  \printtext{in\addspace}
  \printnames{editor}%
  \setunit{\addspace}
  \usebibmacro{booktitle}
  \setunit{\addspace}%
  \printtext{(}\printlist{publisher}%
  \setunit{\normalfont\addcomma\space}%
  \printlist{location}%
  \setunit{\normalfont\addcomma\space}%
  \printfield{year}%
  \printtext{)}%
  \setunit{\normalfont\addcomma\space}%
  \printtext{pp. }\printfield{pages}%
  \usebibmacro{finentry}}

\DeclareBibliographyDriver{inproceedings}{%
  \usebibmacro{bibindex}%
  \usebibmacro{begentry}%
  \normalfont\printnames{author}%
  \setunit{\normalfont\addcomma\space}\newblock
  \usebibmacro{title}
  \newblock
  \printtext{in\addspace}
  \usebibmacro{booktitle}
  \setunit{\addspace}%
  \printtext{(}\printlist{location}%
  \setunit{\normalfont\addcomma\space}%
  \printfield{year}%
  \printtext{)}%
  \setunit{\normalfont\addcomma\space}%
  \printtext{pp. }\printfield{pages}%
  \usebibmacro{finentry}}

\DeclareBibliographyDriver{inbook}{%
  \usebibmacro{bibindex}%
  \usebibmacro{begentry}%
  \normalfont\printnames{author}%
  \setunit{\normalfont\addcomma\space}\newblock
  \usebibmacro{title}
  \newblock
  \printtext{in\addspace}
  \printnames{editor}%
  \setunit{\addspace}
  \usebibmacro{booktitle}
  \setunit{\addspace}%
  \printtext{(}\printlist{publisher}%
  \setunit{\normalfont\addcomma\space}%
  \printlist{location}%
  \setunit{\normalfont\addcomma\space}%
  \printfield{year}%
  \printtext{)}%
  \setunit{\normalfont\addcomma\space}%
  \printtext{pp. }\printfield{pages}%
  \usebibmacro{finentry}}

\DeclareFieldFormat[techreport]{title}{``#1\addcomma''}
\DeclareFieldFormat[techreport]{type}{#1}
\DeclareFieldFormat[techreport]{number}{#1}

\DeclareBibliographyDriver{techreport}{%
  \usebibmacro{bibindex}%
  \usebibmacro{begentry}%
  \normalfont\printnames{author}%
  \setunit{\normalfont\addcomma\space}\newblock
  \usebibmacro{title}
  \newblock
  \printfield{type}
  \setunit{\addspace}%
  \printfield{number}
  \setunit{\normalfont\addcomma\space}%
  \printlist{institution}
  \setunit{\normalfont\addcomma\space}%
  \printlist{location}
  \setunit{\normalfont\addcomma\space}%
  \printfield{year}%
  \usebibmacro{finentry}}

\usepackage{changepage}

\usepackage{algorithm}
\usepackage{algpseudocode}

\begin{document}
\vspace*{0mm}

{\footnotesize{\noindent This is a preprint of the article accepted for publication in {\textit{Mathematical Models and Computer Simulations.}} \textbf{17}, Suppl.\ 2, S263--S278 (2025). © Pleiades Publishing, Ltd., 2025. The final version can be found at \href{https://pleiades.online}{https://pleiades.online}.}}

\begin{center}

\textbf{Multifluid Hydrodynamic Simulation} \\
\vspace*{1mm}
\textbf{of Metallic-Plate Collision Using the VOF Method}
\end{center}

\begin{center}
\fontsize{11}{12.65}\selectfont
{\rm\bfseries
Fedor Belolutskiy$^{a,b,*}$, Elena Oparina$^{a,**}$, Svetlana Fortova$^{a,***}$}\\
\vspace*{3mm}
\normalsize
{\textit{ $^{a}$ICAD RAS, Moscow, Russia} \\
\textit{ $^{b}$Skoltech, Moscow, Russia} \\
\,\textit{$^{*}$e--mail: fedor1113-public@yandex.com} \\
\textit{$^{**}$e--mail: elena\_oparina@mail.ru} \\
\textit{$^{***}$e--mail: sfortova@mail.ru}
} \\
\medskip
%
%
\end{center}

\noindent
{{\textbf{Abstract---}}The present study is concerned with a one-dimensional problem in explosive welding that pertains to the collision of lead and steel plates.
The metal plates and the surrounding air are represented as separate immiscible phases governed by independent equations of state.
A multifluid Godunov-type (finite-volume) computational algorithm, based on the mechanical-equilibrium Euler equations and incorporating pressure relaxation, is used to numerically describe the evolution of the waves resulting from the collision.
The position of the interface (contact discontinuity) between immiscible phases is tracked by means of the volume-of-fluid (VOF) method.
The numerical model allows one to account for the existence of tensile stresses in metal and registers them as regions of negative pressure.
The computed arrival time of the unloading wave at the interface between the plates agrees with the experimental data and with simulation results obtained via different methods.
}

\vspace*{2mm}

\vspace*{1.8mm}

\section{Introduction}

Explosive welding is an established method for joining
dissimilar metals and alloys.
It is a process of metal bonding by way of high-speed impact due to energy released by an explosive.
This process is accompanied by complex dynamic effects, which form the subject matter for many experimental and numerical studies.
Classical scientific research in this field
is associated with the names of A.~A.
Deribas \cite{Deribas:1980en} and S.~K. Godunov \cite{Godunov:1970}.
In O.~M. Belotserkovsky's work \cite{Belotserkovsky:2016en}, it is proven numerically that one of the reasons for the joining of metals may be the Rayleigh--Taylor instability, which arises at the interface of colliding metal plates.

Experimental research conducted for the scientific articles \cite{Yakovlev:1975,Malakhov:2020} shows that welded metals remain in a pseudo-fluid condensed state in the first several microseconds after their collision.
Therefore, the numerical simulation of the collision process may be carried out via a multifluid hydrodynamic approach, based on the solution of the Euler equations.
In this case, every material (phase) is considered a compressible continuum in the Eulerian approximation.

Many fields of science and technology require modelling immiscible compressible continua while accounting for the presence of condensed phases \cite{Nigmatulin:1987book:en}.
Several approaches to the construction of the requisite multiphase models exist in the literature.
Some of them, such as the non-equilibrium multi-velocity %
Rakhmatulin--Nigmatulin model \cite{Rakhmatulin:1956en,Nigmatulin:1987book:en}, %
account for phase transitions and the non-locality of interphase interaction.
Others, like the Baer--Nunziato model \cite{Baer:1986,Chuprov:2021}, %
account for the disequilibrium of the process with respect to velocities, pressures and temperatures, assuming local relaxation.
The homogenised description of compressible media based on the Euler equations with equilibrium pressures and velocities has become widely used \cite{Nigmatulin:1987book:en,Kapila:2001,Povarnitsyn:2006}.
To numerically solve the Euler equation system, %
Godunov-type \cite{Godunov:1976book:en},
discontinuous-Galerkin-type \cite{Hesthaven:2008},
finite-element-type \cite{Donea:2003} and other methods are actively employed.

The main task in numerical simulation of the collision process is to track the phase boundaries (interfaces) of the colliding materials. Various methods \cite{Abgrall:2001, Ho:2023} can be used to capture or track sharp interfaces between immiscible materials, including level-set methods, arbitrary Eulerian--Lagrangian (ALE) methods and volume-of-fluid (VOF) methods.
Among these methodologies, we highlight the VOF method \cite{Hirt:1981} with material-interface reconstruction, which has a number of advantages,
such as relative ease of use, execution speed, global mass conservation and conservation of volume
fraction at the interfacial contact discontinuity. This method is used to track phase interfaces for both incompressible \cite{Hirt:1981} %
and compressible \cite{Cutforth:2021,Law:2024,Menshov:2024en} flows.%

In compressible media, the choice of the method for density and internal-energy determination for each particular phase in computational cells containing several such phases --- so-called \emph{mixed} cells --- presents some difficulty \cite{Norman:1986book}.
Our approach to simulating multiphase flows is based on the idea suggested in an unpublished manuscript by P. Colella, H.~M. Glaz and R.~E. Ferguson (CGF).
In their algorithm, interfaces between phases are tracked using the VOF method with the equations of motion for the multiphase medium complemented by the evolution equations for the volume fractions, energies and densities of each phase in mixed cells. This formulation accommodates using separate and arbitrary equations of state for the phases.
The mathematical model of CGF in the case of two phases is equivalent to Kapila {\textit{et~al.}}'s mechanical-equilibrium five-equation $Vp$-model \cite{Kapila:2001}.
The latter model emerges as a reduction of the non-equilibrium Baer--Nunziato model for describing arbitrary two-phase flows under the assumption that the exchange of momentum and energy at the phase interfaces runs its course at orders-of-magnitude-shorter timescales than advection processes.

In the works \cite{Miller:1996,Cutforth:2021,Law:2024},
the CGF method, originally developed for gases, is improved upon and generalised to condensed-medium problems, subject to various constraints on the utilised equations of state.

In the present work, we suggest a modification of the method due to CGF for multifluid modelling and use different equations of state for each phase.
The developed method is demonstrated on a one-dimensional problem of high-speed collision of metal plates.
It is shown that this approach has low numerical diffusion, thereby yielding higher solution accuracy.
A comparative analysis of the results obtained here with similar computation results from the paper \cite{Chuprov:2021} is performed.

\section{Mathematical model}

\subsection{Effective-phase equations}

We represent the state inside a mixed cell by a single phase with its own density, pressure, internal energy and speed of sound, all consistent with the multiphase continuum.
In accordance with the works \cite{Puckett:1992,Miller:1996,Povarnitsyn:2006,Povarnitsyn:2012en}, this phase is called effective.
Its density is calculated from the conservation of mass together with phase immiscibility via the following formula:
\begin{equation}
\bar{\rho}=\sum_{\alpha=1}^{N_{f}}\rho^{\left(\alpha\right)}f^{\left(\alpha\right)}, \label{eq:effective_density}
\end{equation}
 where $\rho^{\left(\alpha\right)}$ is the actual material density of phase $\left(\alpha\right)$;
$f^{\left(\alpha\right)}$ is the volume fraction (concentration)
of phase $\left(\alpha\right)$ ($\sum_{\alpha=1}^{N_{f}}f^{\left(\alpha\right)}=1$),
while $N_{f}$ is the quantity of phases.

Then the discretised-here effective-phase mathematical model based on the Euler equations in one spatial dimension has the form of the three standard conservation laws applied to that effective phase \cite{Povarnitsyn:2012en}:
\begin{equation}
\partial_{t}\bar{\rho}+\partial_{x}\left(\bar{\rho}u\right)=0,\label{eq:continuity_effective}
\end{equation}
\begin{equation}
\partial_{t}\left(\bar{\rho}u\right)+\partial_{x}\left(\bar{\rho}u^{2}+p\right)=0,
\end{equation}
\begin{equation}
\partial_{t}\left(\bar{\rho}\bar{E}_{\text{total}}\right)+\partial_{x}\left(\bar{\rho}\bar{E}_{\text{total}}u+pu\right)=0,\label{eq:total_energy_effective}
\end{equation}
 where by $u$, $p$, $\bar{E}_{\text{total}}$ and $\bar{\rho}$ we denote the velocity, pressure, total specific energy and density, respectively.

The VOF method tracks the volume fractions of all the phases in a cell.
It is based on the assumption
of a single {\emph{common}} pressure $p$ and a single {\emph{common}} velocity $u$, shared by all phases, which is explained within a mixed cell by the fact that at the interfacial contact discontinuity between phases, the pressure of the continuous
medium and the normal component of velocity do not experience a jump. After any equilibrium-breaking passage of strong waves in the cell, fast pressure equilibration is assumed.

The energy of the effective phase is calculated taking into account that the velocities of all phases in the cell are considered to be the same (i.~e. there is no slip), and the phases are not interpenetrating:
\[
\bar{\rho}\bar{E}_{\text{total}}=\frac{\bar{\rho}u^{2}}{2}+\bar{\rho}\bar{e}=\frac{\bar{\rho}u^{2}}{2}+\sum_{\alpha=1}^{N_{f}}\rho^{\left(\alpha\right)}e^{\left(\alpha\right)}f^{\left(\alpha\right)}=\sum_{\alpha=1}^{N_{f}}\rho^{\left(\alpha\right)}E_{\text{total}}^{\left(\alpha\right)}f^{\left(\alpha\right)},
\]
 where $\bar{e}$ is the specific internal energy of the effective phase, while $e^{\left(\alpha\right)}$
represents the specific internal energy of phase $\left(\alpha\right)$ in accordance with \eqref{eq:stiffened-gas}.

To solve problems in a model with the tracking of interfaces between different phases and the resolution of local Riemann problems in a continuous medium, it is necessary to track the volume fractions of materials in cells in order to determine the coordinates of the interfaces; it is also necessary to separately solve the energy equations and density equations for each phase in order to take into account the different equations of state and compressibilities for these phases.
Hereunder, we present the evolution equations for volume fractions, taking into account compressibility, and the energy equations for individual phases.

The single-velocity model for immiscible inviscid phases considered here is a special case of the model for an $N_f$-phase mixture of viscous compressible phases with common pressure (\cite{Nigmatulin:1987book:en}, pages 15--17).

\subsection{Derivation of the advection equation for the volume fractions of the individual phases accounting for their compressibilities}

To formulate our model, we must derive an advection equation for the volume fractions of the phases.
We start by writing down the continuity equation for each of the phases $\alpha=1,\dots,N_{f}$ with their shared velocity in the absence of mass transfer between them, the absence of phase transitions \cite{Nigmatulin:1987book:en}:
\begin{equation}
\partial_{t}\left(f^{\left(\alpha\right)}\rho^{\left(\alpha\right)}\right)+\partial_{x}\left(f^{\left(\alpha\right)}\rho^{\left(\alpha\right)}u\right)=0.\label{eq:continuity-1}
\end{equation}
 From this, the equation for the volume fractions of the phases follows:
\begin{equation}
\partial_{t}f^{\left(\alpha\right)}+\partial_{x}\left(f^{\left(\alpha\right)}u\right)=-\frac{f^{\left(\alpha\right)}}{\rho^{\left(\alpha\right)}}\partial_{t}\rho^{\left(\alpha\right)}-\frac{f^{\left(\alpha\right)}u}{\rho^{\left(\alpha\right)}}\partial_{x}\rho^{\left(\alpha\right)}.\label{eq:volume_fraction0}
\end{equation}

Now, we need to reformulate the above equation in terms of compression moduli to explicitly include the pressure-equilibrium constraint and ensure that more compressible fluids change their volumes more readily.
Subsequent derivations are performed under the assumption of isentropic compression, i.~e. while holding the entropy of each phase constant during volume-fraction change.
The rationale behind making such an assumption is that materials do not have enough time to redistribute heat among themselves during pressure equalisation, since thermal equilibrium
cannot be achieved in condensed phases under high
pressure on the relevant time and length scales. This assumption should be valid in typical condensed-phase simulations, unless computational cells reach length scales on the order of $ \unit[10^{-9}]{m} $ \cite{Miller:1996,Povarnitsyn:2012en}, which is not the case here.
As such, the relationship between infinitesimal changes of pressure and density is determined by the isentropic volume compression moduli $K_S^{(\alpha)}$.
Recall that the isentropic volume compression modulus, or the isentropic bulk modulus, which we shall from now on always call the \emph{bulk modulus} for brevity, is traditionally defined as a quantity that is the inverse of compressibility and that shows the change in material pressure under uniform volume compression:
\[
K_{S}^{\left(\alpha\right)}\mathrel{\stackrel{\text{def}}{\coloneqq}}-{V^{\left(\alpha\right)}}\left(\partial_{V^{\left(\alpha\right)}}p^{\left(\alpha\right)}\right)_{S}
=\rho^{\left(\alpha\right)}\left(c^{\left(\alpha\right)}\right)^{2}=1/\left(\partial_{p^{\left(\alpha\right)}}\ln\rho^{\left(\alpha\right)}\right)_{S},
\]
 where $S=\text{const}$ denotes the entropy, while $ c^{\left( \alpha \right)} = \sqrt{\left(\partial_{\rho^{\left( \alpha \right)}} p^{\left( \alpha \right)}\right)_S} $
is the sound speed in phase $ \left( \alpha \right) $.

Hence, $ K_{S}^{\left(\alpha\right)}=\rho^{\left(\alpha\right)}\left(\partial_{\rho^{\left( \alpha \right)}} p^{\left( \alpha \right)}\right)_S\sim\rho^{ \left( \alpha \right) }{\delta p^{ \left( \alpha \right) }}/{\delta\rho^{ \left( \alpha \right) }} $, and the infinitesimal change in pressure, $ \delta p^{\left(\alpha\right)} $, along an isentrope may be expressed in terms of the corresponding infinitesimal change in density, $ \delta \rho^{\left(\alpha\right)} $, using the bulk moduli / compressibilities of the materials \cite{Miller:1996}.
For phase $ \left( \alpha \right) $, we have:
\begin{equation}
\delta p^{\left(\alpha\right)} = \frac{K_{S}^{\left(\alpha\right)}}{\rho^{\left(\alpha\right)}}\delta\rho^{\left(\alpha\right)}.\label{eq:isentropic_compression_change}
\end{equation}
In the same way, it is possible to express the infinitesimal isentropic pressure change for the effective phase \cite{Miller:1996}:
\begin{equation}
    \delta\bar{p}=\frac{\bar{K}_{S}}{\bar{\rho}}\delta\bar{\rho}.\label{eq:isentropic_effective_compression_change}
\end{equation}
 The compressibility of the effective phase, $ 1/\bar{K}_{S} $, is defined as a sum of the actual phase compressibilities weighted by the volume fractions of those phases, a formulation that is in agreement with the well-known Wood formula for the sound speeds in a gaseous mixture (\cite{Wood:1930}, pages 327--328):
\[
\bar{K_{S}}=\left(\sum_{\alpha=1}^{N_{f}}\frac{f^{\left(\alpha\right)}}{K_{S}^{\left(\alpha\right)}}\right)^{-1}.
\]

We can arrive at this last definition with the assumption of absent mass-fraction flow between phases in the calculation of instantaneous ``frozen'' compressibility for a material specific volume if we differentiate the following expression with respect to the common pressure:
\[
\sum_{\alpha=1}^{N_{f}}m^{\left( \alpha \right)}\frac{1}{\rho^{\left(\alpha\right)}}=1/\bar{\rho},
\]
 where $ m^{\left( \alpha \right)} = {f^{\left(\alpha\right)}\rho^{\left(\alpha\right)}}/{\bar{\rho}} $ is the mass fraction of phase $ \left( \alpha \right) $.
Differentiating this equality with respect to $ p $ while holding entropies and mass fractions $m^{\left( \alpha \right)}$ constant for all phases, we get:
\[
    -\frac{1}{\bar{\rho}}\sum_{\alpha=1}^{N_{f}}f^{\left(\alpha\right)}\rho^{\left(\alpha\right)}\frac{1}{\rho^{\left(\alpha\right)}}\left(\partial_{p}\ln\rho^{\left(\alpha\right)}\right)_{S}=
\left(\partial_{p}\left(1/{\bar{\rho}}\right)\right)_{S}.
\]
If we combine this with $ \left(\partial_{p^{\left(\alpha\right)}}\ln\rho^{\left(\alpha\right)}\right)_{S} = 1/K_{S}^{\left(\alpha\right)} $, it follows that
\[
    \sum_{\alpha=1}^{N_{f}}\frac{f^{\left(\alpha\right)}}{K_{S}^{\left(\alpha\right)}}=\frac{1}{\bar{\rho}}\left( \partial_{p}\bar{\rho} \right)_S\equiv1/\bar{K}_{S}.
\]

To be able to conclude the derivation and express everything in terms of bulk moduli, thereby excluding densities from the equations, we equate the small change in pressure in \eqref{eq:isentropic_compression_change} to the small change in \eqref{eq:isentropic_effective_compression_change} ($ \delta p^{ \left( \alpha \right) } = \delta \bar{p} = \delta p $), based on the assumption of common pressure and constant entropy of all phases during the redistribution of volume fractions in a mixed cell \cite{Miller:1996}. The relationship holds for small changes along the trajectory of a particle in a single-velocity continuous medium. Consequently, for total (Lagrangian) derivatives $ d/dt = \partial_t + u\partial_x $ of densities corresponding to their small changes, the following is true:
\[
    \frac{1}{ \rho^{ \left( \alpha \right) } } \frac {d\rho^{ \left( \alpha \right) }}{dt} = \frac{\bar{K}_S}{ K_S^{ \left( \alpha \right) } } \frac{1}{\bar{\rho}} \frac{d \bar{\rho}}{dt}.
\]
In that case, we can rearrange equation \eqref{eq:volume_fraction0} in the following form:
\[
    \partial_{t}f^{\left(\alpha\right)}+\partial_{x}\left(f^{\left(\alpha\right)}u\right)
=-\frac{f^{\left(\alpha\right)}}{\rho^{\left(\alpha\right)}} \frac{d \rho^{ \left( \alpha \right) }}{dt},
\]
\[
\partial_{t}f^{\left(\alpha\right)}+\partial_{x}\left(f^{\left(\alpha\right)}u\right)=-\frac{f^{\left(\alpha\right)}\bar{K}_{S}}{K_{S}^{\left(\alpha\right)}}\frac{1}{\bar{\rho}}\frac{d \bar{\rho}}{dt}.
\]
Finally, substituting $ d \bar{\rho}/dt $ from the effective-phase continuity equation (equation \eqref{eq:continuity_effective}) into the last equation, since $ \partial_t \bar{\rho} + u \partial_x  \bar{\rho} + \bar{\rho} \partial_x u = 0 $, we get:
\[
\partial_{t}f^{\left(\alpha\right)}+\partial_{x}\left(f^{\left(\alpha\right)}u\right)=-\frac{f^{\left(\alpha\right)}\bar{K}_{S}}{\bar{\rho}K_{S}^{\left(\alpha\right)}}\left(\partial_{t}\bar{\rho}+u\partial_{x}\bar{\rho}\right),
\]
\begin{equation}
\partial_{t}f^{\left(\alpha\right)}+\partial_{x}\left(f^{\left(\alpha\right)}u\right)=\frac{f^{\left(\alpha\right)}\bar{K}_{S}}{K_{S}^{\left(\alpha\right)}}\partial_{x}u,\qquad\alpha=1,\dots,N_{f}.\label{eq:volume_fraction_equation}
\end{equation}
This is the sought-after equation for the advection of the phase volume fractions (l.h.s.) accounting for the compressibilities of the phases (r.h.s.).
The contribution of a phase towards the change in local volume is proportional to the phase's own compressibility $1/K_S^{(\alpha)}$ relative to the compressibility of the mixture as a whole, $1/\bar{K}_S$. According to \eqref{eq:volume_fraction_equation},
phases
with greater compressibilities $1/K_{S}^{\left(\alpha\right)}$ will be more strongly
compressed by the total divergence of velocity.

\subsection{Derivation of the energy equation for the individual phases}

Let us now derive the energy equation for individual phases.
In the inviscid model under consideration, there is no interaction between phases, not only in terms of mass, but also in terms of force (due to interphase friction, surface tension etc.) and energy (due to the work of interphase forces, heat transfer etc.).
We do not consider the effect of external force fields on a multiphase medium, either.
In this case, the internal energy equation of the model with a common pressure from \cite{Nigmatulin:1987book:en} degenerates into the following equation, which coincides with the single-phase case:
\[
    \rho^{\left(\alpha\right)} \frac{d}{dt} e^{\left(\alpha\right)} = \frac{p}{\rho^{\left(\alpha\right)}}\frac{d\rho^{\left(\alpha\right)}}{dt}.
\]
 In the energy equation in a multiphase medium, compared to a single-phase flow, the additional effect of different phase compressibilities needs to be addressed, consistently with \eqref{eq:volume_fraction_equation}.
For a single-phase flow away from discontinuities, the internal energy equation can be written in terms of the volume of the phase:
\begin{equation}
\rho^{\left(\alpha\right)}\frac{d}{dt}e^{\left(\alpha\right)}+\frac{p^{\left(\alpha\right)}}{V^{\left(\alpha\right)}}\frac{dV^{\left(\alpha\right)}}{dt}=0. \label{internal_energy_eq_via_vol_change}
\end{equation}
In a multiphase flow, the following relationship must be satisfied for infinitesimal changes in the volume of each phase, obtained similarly to the reasoning for equations \eqref{eq:isentropic_compression_change} and \eqref{eq:isentropic_effective_compression_change}:
\begin{equation}
\frac{1}{V^{\left(\alpha\right)}} \frac{d V^{\left(\alpha\right)}}{dt} =\frac{\bar{K}_{S}}{K_{S}^{\left(\alpha\right)}}\frac{1}{\bar{V}} \frac{d\bar{V}}{dt}. \label{eq:isentropic_compression_vol_change}
\end{equation}

Substituting \eqref{eq:isentropic_compression_vol_change} into equation \eqref{internal_energy_eq_via_vol_change}, taking into account the equality $p^{\left(\alpha\right)} = p$ and the equation $ \bar{V}^{-1}d\bar{V}/dt = \partial_x u $ for the dilatation rate, leads us to the following form of the equation for internal energy:
\begin{equation}
\rho^{\left(\alpha\right)}\frac{d}{dt}e^{\left(\alpha\right)}+p\frac{\bar{K}_{S}}{K_{S}^{\left(\alpha\right)}}\partial_{x}u=0.\label{eq:internal_energy_law}
\end{equation}
This form of the equation takes into account that phases in contact cannot compress and expand independently.

At last, we obtain the equation for the total specific energy of one phase, $ E^{ \left( \alpha \right) }_\text{total} = \left( u^{\left( \alpha \right)}\right)^2/2 + e^{\left( \alpha \right)} $, summing equation \eqref{eq:internal_energy_law} with the momentum conservation equation multiplied by $ \rho^{\left(\alpha\right)}u $, \[\rho^{\left(\alpha\right)}u\,\frac{du}{dt}=-\frac{\rho^{\left(\alpha\right)}u}{\bar{\rho}}\partial_{x}p,\] and multiplying the result by $f^{\left(\alpha\right)}$:
\[
f^{\left(\alpha\right)}\rho^{\left(\alpha\right)}\frac{d}{dt}E_{\text{total}}^{\left(\alpha\right)}+\frac{f^{\left(\alpha\right)}\rho^{\left(\alpha\right)}u}{\bar{\rho}}\partial_{x}p=-p\frac{f^{\left(\alpha\right)}\bar{K}_{S}}{K_{S}^{\left(\alpha\right)}}\partial_{x}u,
\]
\begin{equation}
\partial_{t}\left(f^{\left(\alpha\right)}\rho^{\left(\alpha\right)}E_{\text{total}}^{\left(\alpha\right)}\right)+\partial_{x}\left(f^{\left(\alpha\right)}\rho^{\left(\alpha\right)}E_{\text{total}}^{\left(\alpha\right)}u\right)+u\frac{f^{\left(\alpha\right)}\rho^{\left(\alpha\right)}}{\bar{\rho}}\partial_{x}p=-p\frac{f^{\left(\alpha\right)}\bar{K}_{S}}{K_{S}^{\left(\alpha\right)}}\partial_{x}u.\label{eq:total_energy_law}
\end{equation}

Equation \eqref{eq:total_energy_law} is used for cells containing only one phase.

In mixed cells, equation \eqref{eq:internal_energy_law} multiplied by $ f^{ \left( \alpha \right) } $ is used:
\begin{equation}
\partial_{t}\left(f^{\left(\alpha\right)}\rho^{\left(\alpha\right)}e^{\left(\alpha\right)}\right)+\partial_{x}\left(f^{\left(\alpha\right)}\rho^{\left(\alpha\right)}e^{\left(\alpha\right)}u\right)=-p\frac{f^{\left(\alpha\right)}\bar{K}_{S}}{K_{S}^{\left(\alpha\right)}}\partial_{x}u\label{eq:internal_energy}
\end{equation}
 together with
 $ E^{ \left( \alpha \right) }_\text{total} = \left( u^{\left( \alpha \right)}\right)^2/2 + e^{\left( \alpha \right)}, $
 since using this equation for internal
energies increases
the robustness of numerical schemes \cite{Cutforth:2021}.
The right-hand side of equation \eqref{eq:internal_energy} corresponds to the work done by pressure
on the expansion/compression of phase $ \left( \alpha \right) $, $pdV^{\left(\alpha\right)}$.

\subsection{Final form of the mathematical model and its properties}

The final model is as follows:
\begin{equation}
\partial_{t}f^{\left(\alpha\right)}+\partial_{x}\left(f^{\left(\alpha\right)}u\right)=\frac{f^{\left(\alpha\right)}\bar{K}_{S}}{K_{S}^{\left(\alpha\right)}}\partial_{x}u,\qquad\alpha=1,\dots,N_{f},\label{eq:volume_fraction_equation-1}
\end{equation}
\begin{equation}
\partial_{t}\left(f^{\left(\alpha\right)}\rho^{\left(\alpha\right)}\right)+\partial_{x}\left(f^{\left(\alpha\right)}\rho^{\left(\alpha\right)}u\right)=0,\qquad\alpha=1,\dots,N_{f},\label{eq:continuity-2}
\end{equation}
\begin{equation}
\partial_{t}\left(\bar{\rho}u\right)+\partial_{x}\left(\bar{\rho}u^{2}+p\right)=0\label{eq:momentum}
\end{equation}
 and
\begin{equation}
\partial_{t}\left(f^{\left(\alpha\right)}\rho^{\left(\alpha\right)}E_{\text{total}}^{\left(\alpha\right)}\right)+\partial_{x}\left(f^{\left(\alpha\right)}\rho^{\left(\alpha\right)}E_{\text{total}}^{\left(\alpha\right)}u\right)+u\frac{f^{\left(\alpha\right)}\rho^{\left(\alpha\right)}}{\bar{\rho}}\partial_{x}p=-p\frac{f^{\left(\alpha\right)}\bar{K}_{S}}{K_{S}^{\left(\alpha\right)}}\partial_{x}u,\qquad\alpha=1,\dots,N_{f},\label{eq:total_energy}
\end{equation}
 where the system can be closed by arbitrary caloric equations of state,
 \[\left(\rho^{\left(\alpha\right)},e^{\left(\alpha\right)}\right)\stackrel{\text{EoS }\left(\alpha\right)}{\mapsto}p^{\left(\alpha\right)},\qquad\alpha=1,\dots,N_{f},\]
 but pressures $p^{\left(\alpha\right)}$ relax towards the common pressure $p$ (i.~e. $p^{\left(\alpha\right)}=p$), as facilitated by the fast pressure equilibration assumption.

In the problem under consideration, $ N_f = 3 $, and a stiffened-gas-type \cite{Godunov:1976book:en} equation of state is used for each phase:
\begin{equation}
p^{\left(\alpha\right)}\left(\rho^{\left(\alpha\right)},e^{\left(\alpha\right)}\right)=\left(\gamma^{\left(\alpha\right)}-1\right)\rho^{\left(\alpha\right)} e^{\left(\alpha\right)}-\gamma^{\left(\alpha\right)} p_{\infty}^{\left(\alpha\right)},\qquad\alpha=1,\dots,N_{f},\label{eq:stiffened-gas}
\end{equation}
 where $p^{\left(\alpha\right)}$, $\rho^{\left(\alpha\right)}$, $e^{\left(\alpha\right)}$ refer to the pressure, density and energy in phase $ \left(\alpha\right) $, while $\gamma^{\left(\alpha\right)}$ and $p_{\infty}^{\left(\alpha\right)}$ are the parameters of its equation of state.
The speed of sound in phase $ \left( \alpha \right) $ is then calculated using the formula:
\begin{equation}
c^{\left(\alpha\right)}=\sqrt{\gamma^{\left(\alpha\right)} \left( p^{\left(\alpha\right)}+p^{\left(\alpha\right)}_{\infty} \right) / \rho^{\left(\alpha\right)}}.
\label{eq:sound_speed}
\end{equation}

Summing all phases $ \alpha = 1, \dots, N_f $ from \eqref{eq:continuity-2}, we get equation \eqref{eq:continuity_effective}. In the same way, equation  \eqref{eq:total_energy_effective} is obtained
from \eqref{eq:total_energy}.
Far from phase boundaries, the equations of this model degenerate
into the standard Euler equations for a single phase.

The form of equations \eqref{eq:continuity-2}--\eqref{eq:momentum} for a single phase is conservative, and the form of $\eqref{eq:total_energy}$ is close to conservative, by which we mean that $u\partial_{x}p^{\left(\alpha\right)}+p^{\left(\alpha\right)}\partial_{x}u$
admits a natural conservative discretisation: $\tilde{u}_{i}\left(p_{i+1/2}^{n+1/2}-p_{i-1/2}^{n+1/2}\right)/{\Delta x}+\tilde{p}_{i}\left(u_{i+1/2}^{n+1/2}-u_{i-1/2}^{n+1/2}\right)/{\Delta x}=$
\begin{multline}
\frac{1}{2}\left(u_{i+1/2}^{n+1/2}+u_{i-1/2}^{n+1/2}\right)\frac{1}{\Delta x}\left(p_{i+1/2}^{n+1/2}-p_{i-1/2}^{n+1/2}\right)+\frac{1}{2}\left(p_{i+1/2}^{n+1/2}+p_{i-1/2}^{n+1/2}\right)\frac{1}{\Delta x}\left(u_{i+1/2}^{n+1/2}-u_{i-1/2}^{n+1/2}\right)\\
=\frac{1}{\Delta x}\left(u_{i+1/2}^{n+1/2}p_{i+1/2}^{n+1/2}-p_{i-1/2}^{n+1/2}u_{i-1/2}^{n+1/2}\right),
\end{multline}
 which coincides with the discretisation of $\partial_{x}\left(pu\right)$ from
the conservative form of the energy equation.

Let us list the properties of system \eqref{eq:volume_fraction_equation-1}--\eqref{eq:total_energy} \cite{Puckett:1992}.
Equations \eqref{eq:volume_fraction_equation-1}
preserve $\sum_{\alpha=1}^{N_{f}}f^{\left(\alpha\right)}=1$. All the volume fractions $ f^{\left(\alpha\right)} $ individually remain between zero and one if originally in that range.
System \eqref{eq:volume_fraction_equation-1}--\eqref{eq:total_energy}
is hyperbolic if and only if
the speed of sound of the effective phase is real, i.e., the ratio of the effective phase's bulk modulus to its density is greater than or equal to zero: $\bar{K}_{S}/\bar{\rho}\geqslant0$. System \eqref{eq:volume_fraction_equation-1}--\eqref{eq:total_energy}
maintains the equilibrium of phase pressures away from discontinuities. 

Thus, the presented system of equations (\eqref{eq:volume_fraction_equation-1}--\eqref{eq:total_energy}) is hyperbolic and thermodynamically consistent \cite{Miller:1996}.

\section{Problem statement}

\begin{figure}
\begin{center}
{\large\textbf{
\includegraphics[scale=0.6]{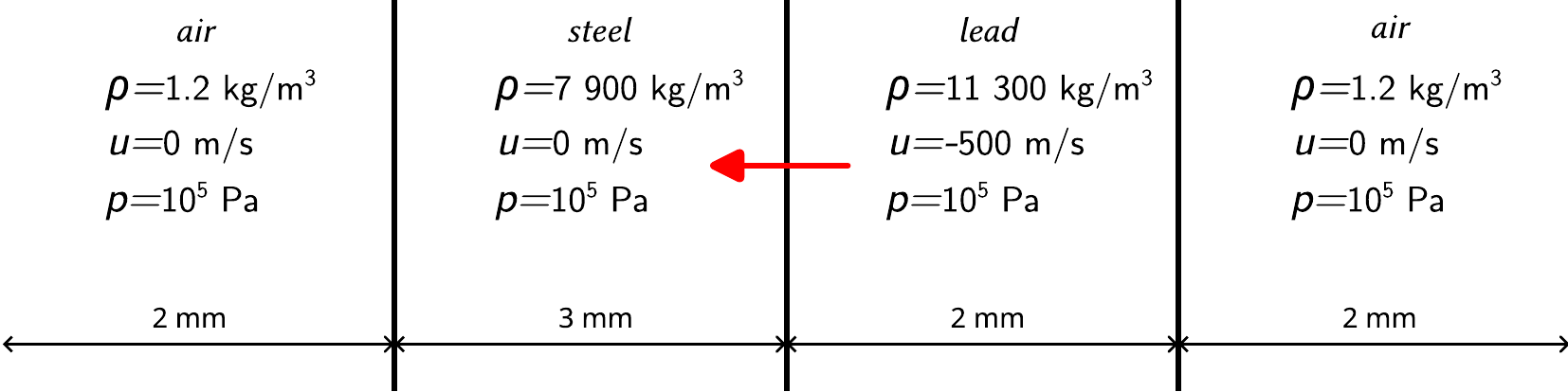}
}}{\large\par}
\par\end{center}
\starcaption{
A schematic for the distribution of densities, velocities and pressures at time $ t = 0 $.
The left-pointing arrow indicates the direction of movement of the lead plate towards the steel plate.
}\label{fig:1}
\end{figure}

Consider a high-speed collision between a lead plate (phase 1), having a density of $ \unit[11\,300]{kg/m^3} $ and a thickness of $ \unit[2]{mm} $, and a steel plate (phase 2), having a density of $ \unit[7\,900]{kg/m^3} $ and a thickness of $ \unit[3]{mm} $. The velocity of the lead plate relative to the steel plate is $ \unit[500]{m/s} $. The metal plates are surrounded by a layer of air (phase 3) with a density of $ \unit[1.2]{kg/m^3} $ and a thickness of $ \unit[2]{mm} $ (Fig.\ \ref{fig:1})
The initial pressure is uniform throughout the entire region and is equal to $ \unit[10^5]{Pa} $.

 The computational domain is the segment $ \left[ \unit[0]{mm} \mathop{..} \unit[9]{mm} \right] $ with impermeable-wall boundary conditions. The colliding surfaces remain in a pseudo-fluid state for a period on the order of several microseconds. Computations are performed up to $ t = \unit[2]{\mus} $, a duration which exceeds the time it takes for the unloading wave from the lead plate to reach the contact interface with the steel plate.

All phases are modelled with equations of state of the stiffened-gas type \eqref{eq:stiffened-gas}. The values of the parameters of these equations are presented in \cite{Utkin:2018conf} and are as follows: for steel, $p_{\infty}=\unit[65]{\text{GPa}}$, $\gamma=3$; for lead, $p_{\infty}=\unit[15.5]{\text{GPa}}$, $\gamma=2.7$; for air, $p_{\infty}=\unit[0]{\text{GPa}}$, $\gamma=1.4$.

The problem statement corresponds to a full-scale physical experiment from \cite{Yakovlev:1975}.

\section{Numerical method}

\begin{figure}
\begin{center}
{\large\textbf{
  \includegraphics[scale=0.7]{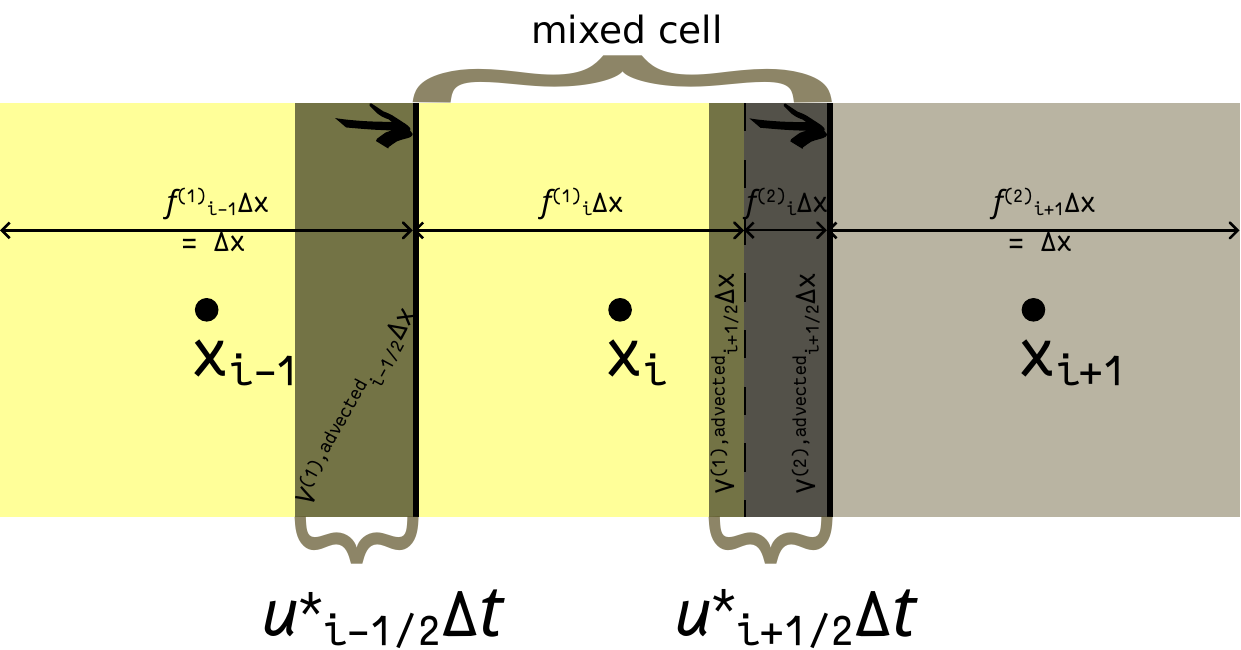}
}}{\large\par}
\par\end{center}
\starcaption{
Advection of Lagrangian volumes (drawn as thin shaded rectangles) through a mixed
cell (a cell centred at $ x_i $) for two phases in a constant  velocity field directed to the right. The phase interface
is located at the point $x_{i-1} + 0.5 \Delta x + f_{i}^{\left(1\right)}\Delta x$ inside
the Eulerian cell, of size $\Delta x$ with coordinate $x_{i}$
}\label{fig:2}
\end{figure}

The VOF method \cite{Hirt:1981} relies on an Eulerian approach to the description of a continuous medium with volume-fraction advection for the phases.
A standard step in the VOF method consists of the computation of the volumes flowing in and out of all Eulerian cells, as dictated by the left-hand side of equation \eqref{eq:volume_fraction_equation}.
Over a time step of $ \Delta t $, a volume equal to $ V^\text{advected} = \Delta t \, u^{\ast} $ flows across the cell interface along the $ x $-axis.
The velocities
$u_{i-1/2}^{\ast}$, $u_{i+1/2}^{\ast}$ at the cell interfaces are found from the solution of the local Riemann problems between
the effective phases in the cells sharing those interfaces.
In the one-dimensional case,
the calculation of the volumes of the phases flowing between cells corresponds to the donor-acceptor scheme \cite{Ramshaw:1976}:
\begin{equation}
\tilde{f}_{i}^{\left(\alpha\right),n+1}\leftarrow f_{i}^{\left(\alpha\right),n}-\left(V_{i+1/2}^{\left(\alpha\right),\text{advected}}-V_{i-1/2}^{\left(\alpha\right),\text{advected}}\right),\qquad\alpha=1,\dots,N_{f},\label{eq:vof_upd}
\end{equation}
 where, e.~g., for a flow of $ N_f=2 $ phases directed to the right, depicted in
Fig.\ \ref{fig:2}, $u_{i-1/2}^{\ast},u_{i+1/2}^{\ast}>0$,
\[
u_{i+1/2}^{\ast}\Delta t=\Delta x\left(\left(V_{i+1/2}^{\left(1\right),\text{advected}}\right)+\left(V_{i+1/2}^{\left(2\right),\text{advected}}\right)\right)=\Delta x\left(\left(\left|\frac{u_{i+1/2}^{\ast}\Delta t}{\Delta x}\right|-\left(1-f_{i}^{\left(1\right)}\right)\right)+\left(f_{i}^{\left(2\right)}\right)\right)
\]
 (note that $u_{i-1/2}^{\ast}\Delta t=\Delta xV_{i-1/2}^{\left(1\right),\text{advected}}$;
$f_{i}^{\left(1\right)}+f_{i}^{\left(2\right)}=1$).
\enlargethispage{1\baselineskip}

Using $\tilde{f}^{(\alpha)}$, the effective compressibility $1/\tilde{K}_{S}^{(\alpha)}$ of the influent material is found by first calculating the volume-fraction-weighted average of individual compressibilities --- both already present in the cell and currently inflowing --- and then dividing this average by $\tilde{f}^{(\alpha)}$. After that, the actual volume fraction $f^{\left(\alpha\right)}$ is updated according to the following formula:
\vspace{-0.2\baselineskip}
\begin{equation}
f_{i}^{\left(\alpha\right),n+1}\leftarrow\tilde{f}_{i}^{\left(\alpha\right),n+1}+\frac{\tilde{f}_{i}^{\left(\alpha\right),n+1}\left(\sum_{\alpha=1}^{N_{f}}\tilde{f}_{i}^{\left(\alpha\right),n+1}/\left(\tilde{K}_{S}^{\left(\alpha\right)}\right)_{i}^{n+1}\right)^{-1}}{\left(\tilde{K}_{S}^{\left(\alpha\right)}\right)_{i}^{n+1}}\overbrace{\left(1-\sum_{\beta=1}^{N_{f}}\tilde{f}_{i}^{\left(\beta\right),n+1}\right)}^{\smash{\substack{\Delta t \, (\partial_{x}u)_i\\
\shortparallel\\
\qquad
}
}}.\label{eq:compressibility_correction_upd}
\end{equation}

Let us now outline the general algorithm. Consider a computational domain with a regular Eulerian grid
having a constant step size $\Delta x$ and partitioned into $N_{x}$ cells. Let us
assume that there are $N_{f}$ different phases (materials) in the problem.
In each $i$th cell at the $n$th time step, we assign the volume
fractions $f_{i}^{\left(\alpha\right),n}$, densities $\rho_{i}^{\left(\alpha\right),n}$,
total specific energies $E_{i}^{\left(\alpha\right),n}$, pressures
$p_{i}^{\left(\alpha\right),n}$, and bulk moduli $K_{Si}^{\left(\alpha\right),n}$
to the corresponding materials $\alpha=1,\dots,N_{f}$.
Each cell also has the following effective-phase variables associated with it: effective density $\bar{\rho}_{i}^{n}$, pressure $\bar{p}_{i}^{n}$, velocity $ u_{i}^{n} $, total energy $\bar{\rho E}_{i}^{n}$ and bulk modulus $\bar{K_{S}}_{i}^{n}$.
These effective variables are calculated from the variables of the real phases actually present in the cell.
Then the algorithm corresponding to one step in time is as follows.
\begin{enumerate}
\item Compute the effective variables in
the interior domain:
\[
\bar{\rho}_{i}^{n}\leftarrow\sum_{\alpha=1}^{N_{f}}f_{i}^{\left(\alpha\right),n}\rho_{i}^{\left(\alpha\right),n},\qquad i=0,\dots,N_{x}-1,
\]
\[
\bar{\rho E}_{i}^{n}\leftarrow\frac{\bar{\rho}_{i}^{n}}{2}\left(u_{i}^{n}\right)^{2}+\sum_{\alpha=1}^{N_{f}}f_{i}^{\left(\alpha\right),n}\rho_{i}^{\left(\alpha\right),n}\left(E_{i}^{\left(\alpha\right),n}-\frac{1}{2}\left(u_{i}^{n}\right)^{2}\right),\qquad i=0,\dots,N_{x}-1,
\]
\[
\bar{K_{S}}_{i}^{n}\leftarrow\left(\sum_{\alpha=1}^{N_{f}}\frac{f_{i}^{\left(\alpha\right),n}}{\left(K_{S}^{\left(\alpha\right)}\right)_{i}^{n}}\right)^{-1},\qquad i=0,\dots,N_{x}-1,
\]
\[
\bar{p}_{i}^{n}\leftarrow\sum_{\alpha=1}^{N_{f}}\frac{f_{i}^{\left(\alpha\right),n}p_{i}^{\left(\alpha\right),n}}{\left(K_{S}^{\left(\alpha\right)}\right)_{i}^{n}}/\sum_{\alpha=1}^{N_{f}}\frac{f_{i}^{\left(\alpha\right),n}}{\left(K_{S}^{\left(\alpha\right)}\right)_{i}^{n}}=\left(\bar{K_{S}}\right)_{i}^{n}\sum_{\alpha=1}^{N_{f}}\frac{f_{i}^{\left(\alpha\right),n}p_{i}^{\left(\alpha\right),n}}{\left(K_{S}^{\left(\alpha\right)}\right)_{i}^{n}},\qquad i=0,\dots,N_{x}-1.
\]
\item Apply boundary conditions by defining and populating $N_{g}$ ghost cells at each end of the domain with indices $i=-1,\dots,-N_{g}$
and $N_{x},\dots,N_{x}+N_{g}-1$ according to the prescribed type.

\item Reconstruct the cell-interface effective variables:
$\bar{\rho}_{i+1/2}^{\ell}$, $u_{i+1/2}^{\ell}$, $\bar{p}_{i+1/2}^{\ell}$, $\bar{\rho E}_{i+1/2}^{\ell}$, $\bar{K_{S}}_{i+1/2}^{\ell}$ at the left cell interfaces and $\bar{\rho}_{i+1/2}^{r}$,
$u_{i+1/2}^{r}$, $\bar{p}_{i+1/2}^{r}$, $\bar{\rho E}_{i+1/2}^{r}$,
$\bar{K_{S}}_{i+1/2}^{r}$ at the right
($i=0,\dots,N_{x}-1$).

\item Calculate the time step $ \Delta t $ complying with the Courant--Friedrichs--Lewy condition for
the effective phase
($\text{CFL}=0.1$ in our problems unless stated otherwise):
\[
\Delta t\leftarrow \text{CFL}\cdot\Delta x / \mathop{\underset{i=0,\dots,N_{x}-1}{\max}}\left\{ \left|u_{i}^{n}\right|+\sqrt{\bar{K_{S}}_{i}^{n}/\bar{\rho}_{i}^{n}}\right\}.
\]

\item Solve the local Riemann problems at cell interfaces using the HLLC solver with the reconstructed effective variables to find velocities $u_{i+1/2}^{n+1/2}$ and pressures $\bar{p}_{i+1/2}^{n+1/2}$ ($i=0,\dots,N_{x}-1$).
\item
Determine the phase volumes and compressibilities flowing into each cell ($i=0,\dots,N_{x}-1$):
\begin{enumerate}
\item compute advected volume fractions according to the donor-acceptor scheme via \eqref{eq:vof_upd};
\item find the average effective compressibility of phases $\alpha=1,\dots,N_{f}$,
of the material flowing into the cell through
\[
\left(\left(\tilde{K}_{S}^{\left(\alpha\right)}\right)_{i}^{n+1}\right)^{-1}\leftarrow\left(\tilde{f}_{i}^{\left(\alpha\right),n+1}\right)^{-1}\begin{cases}
\frac{f_{i}^{\left(\alpha\right),n}}{\left(K_{S}^{\left(\alpha\right)}\right)_{i}^{n}}+\frac{V_{i-1/2}^{\left(\alpha\right),\text{advected}}}{\left(K_{S}^{\left(\alpha\right)}\right)_{i-1}^{n}}+\frac{V_{i+1/2}^{\left(\alpha\right),\text{advected}}}{\left(K_{S}^{\left(\alpha\right)}\right)_{i+1}^{n}}, & u_{i-1/2}^{n+1/2}>0\text{ and }u_{i+1/2}^{n+1/2}<0\\
\frac{f_{i}^{\left(\alpha\right),n}+V_{i-1/2}^{\left(\alpha\right),\text{advected}}}{\left(K_{S}^{\left(\alpha\right)}\right)_{i}^{n}}+\frac{V_{i+1/2}^{\left(\alpha\right),\text{advected}}}{\left(K_{S}^{\left(\alpha\right)}\right)_{i+1}^{n}}, & u_{i-1/2}^{n+1/2}\leqslant0\text{ and }u_{i+1/2}^{n+1/2}<0\\
\frac{f_{i}^{\left(\alpha\right),n}-V_{i+1/2}^{\left(\alpha\right),\text{advected}}}{\left(K_{S}^{\left(\alpha\right)}\right)_{i}^{n}}+\frac{V_{i-1/2}^{\left(\alpha\right),\text{advected}}}{\left(K_{S}^{\left(\alpha\right)}\right)_{i-1}^{n}}, & u_{i-1/2}^{n+1/2}>0\text{ and }u_{i+1/2}^{n+1/2}\geqslant0\\
\frac{f_{i}^{\left(\alpha\right),n}-V_{i+1/2}^{\left(\alpha\right),\text{advected}}+V_{i-1/2}^{\left(\alpha\right),\text{advected}}}{\left(K_{S}^{\left(\alpha\right)}\right)_{i}^{n}}, & u_{i-1/2}^{n+1/2}\leqslant0\text{ and }u_{i+1/2}^{n+1/2}\geqslant0;
\end{cases}
\]
\item compute new, corrected volume fractions $ f^{\left( \alpha \right),n+1}_i $ of the phases 
in the cells via \eqref{eq:compressibility_correction_upd}.
\end{enumerate}
\item Compute the fluxes and auxiliary variables at cell interfaces. For all phases $\alpha=1,\dots,N_{f}$ and all cell interfaces $i-1/2$ (where $i=0,\dots,N_{x}$), calculate the advected values of the densities and energies of these phases using the upwind values of the variables:
\[
\tilde{\rho}_{i-1/2}^{\left(\alpha\right),n+1/2}\leftarrow\begin{cases}
\rho_{i-1}^{\left(\alpha\right),n}, & u_{i-1/2}^{n+1/2}>0\\
\rho_{i}^{\left(\alpha\right),n}, & u_{i-1/2}^{n+1/2}\leqslant0,
\end{cases}
\]
\[
\tilde{\rho}_{i-1/2}^{n+1/2}\coloneqq\frac{\sum_{ \left( \alpha \right) }V_{i-1/2}^{\left(\alpha\right),\text{advected}}\tilde{\rho}_{i-1/2}^{\left(\alpha\right),n+1/2}}{\sum_{ \left( \alpha \right) }V_{i-1/2}^{\left(\alpha\right),\text{advected}}},
\]
\[
\tilde{E}_{i-1/2}^{\left(\alpha\right),n+1/2}\leftarrow\begin{cases}
E_{i-1}^{\left(\alpha\right),n}-\frac{1}{2}\left(u_{i-1}^{n}\right)^{2}+\frac{1}{2}\left(u_{i-1/2}^{n+1/2}\right)^{2}, & u_{i-1/2}^{n+1/2}>0\\
E_{i}^{\left(\alpha\right),n}-\frac{1}{2}\left(u_{i}^{n}\right)^{2}+\frac{1}{2}\left(u_{i-1/2}^{n+1/2}\right)^{2}, & u_{i-1/2}^{n+1/2}\leqslant0,
\end{cases}
\]
\[
\tilde{e}_{i-1/2}^{\left(\alpha\right),n+1/2}\leftarrow\begin{cases}
E_{i-1}^{\left(\alpha\right),n}-\frac{1}{2}\left(u_{i-1}^{n}\right)^{2}, & u_{i-1/2}^{n+1/2}>0\\
E_{i}^{\left(\alpha\right),n}-\frac{1}{2}\left(u_{i}^{n}\right)^{2}, & u_{i-1/2}^{n+1/2}\leqslant0.
\end{cases}
\]
Upwind velocity values:
\[
\tilde{u}_{i-1/2}^{n+1/2}\leftarrow\begin{cases}
u_{i-1}^{n}, & u_{i-1/2}^{n+1/2}>0\\
u_{i}^{n}, & u_{i-1/2}^{n+1/2}\leqslant0.
\end{cases}
\]
Simple averages (arithmetic means) of velocities and pressures:
\[
u_{i}^{n+1/2}\leftarrow\frac{1}{2}\left(u_{i-1/2}^{n+1/2}+u_{i+1/2}^{n+1/2}\right),
\]
\[
\bar{p}_{i}^{n+1/2}\leftarrow\frac{1}{2}\left(\bar{p}_{i-1/2}^{n+1/2}+\bar{p}_{i+1/2}^{n+1/2}\right).
\]
\item Compute new density values for all phases $\alpha=1,\dots,N_{f}$ to compute the density of the effective phase at the next time step ($i=0,\dots,N_{x}-1$):
\[
\rho_{i}^{\left(\alpha\right),n+1}\leftarrow\left(f_{i}^{\left(\alpha\right),n+1}\right)^{-1}\left(f_{i}^{\left(\alpha\right),n}\rho_{i}^{\left(\alpha\right),n}-\left(V_{i+1/2}^{\left(\alpha\right),\text{advected}}\tilde{\rho}_{i+1/2}^{\left(\alpha\right),n+1/2}-V_{i-1/2}^{\left(\alpha\right),\text{advected}}\tilde{\rho}_{i-1/2}^{\left(\alpha\right),n+1/2}\right)\right),
\]
\[
\overline{\rho}_{i}^{n+1}\mathrel{\stackrel{}{\leftarrow}}\sum_{\alpha=1}^{N_{f}}f_{i}^{\left(\alpha\right),n+1}\rho_{i}^{\left(\alpha\right),n+1}.
\]
\item Compute the velocity value at the next time step ($i=0,\dots,N_{x}-1$):
\[
u_{i}^{n+1}\leftarrow\left(\overline{\rho}_{i}^{n+1}\right)^{-1}\left(\overline{\rho}_{i}^{n}u_{i}^{n}-\frac{\Delta t}{\Delta x}\left(\left(\tilde{\rho}_{i+1/2}^{n+1/2}\left(\tilde{u}_{i+1/2}^{n+1/2}\right)^{2}+\bar{p}_{i+1/2}^{n+1/2}\right)-\left(\tilde{\rho}_{i-1/2}^{n+1/2}\left(\tilde{u}_{i-1/2}^{n+1/2}\right)^{2}+\bar{p}_{i-1/2}^{n+1/2}\right)\right)\right).
\]
\item Compute the total energy values at the next time step ($i=0,\dots,N_{x}-1$). In cells that are multiphase, adjacent to multiphase or bordering those adjacent cells,
update the internal energy $e_{i}^{\left(\alpha\right),n+1}$
inside the total using
\begin{multline}
E_{i}^{\left(\alpha\right),n+1}\leftarrow\left(f_{i}^{\left(\alpha\right),n+1}\rho_{i}^{\left(\alpha\right),n+1}\right)^{-1}f_{i}^{\left(\alpha\right),n}\overline{\rho}_{i}^{n} \left( E_{i}^{\left(\alpha\right),n} - \frac{1}{2} \left( u_i^n \right)^2 \right) \\
-\Biggl(V_{i+1/2}^{\left(\alpha\right),\text{advected}}\tilde{\rho}_{i+1/2}^{\left(\alpha\right),n+1/2}\tilde{e}_{i+1/2}^{\left(\alpha\right),n+1/2}  -  V_{i-1/2}^{\left(\alpha\right),\text{advected}}\tilde{\rho}_{i-1/2}^{\left(\alpha\right),n+1/2}\tilde{e}_{i-1/2}^{\left(\alpha\right),n+1/2}\Biggr)\left(f_{i}^{\left(\alpha\right),n+1}\rho_{i}^{\left(\alpha\right),n+1}\right)^{-1}\\
-\tilde{p}_{i}^{\left(\alpha\right)}\left(f_{i}^{\left(\alpha\right),n+1}-\tilde{f}_{i}^{\left(\alpha\right),n+1}\right)\left(f_{i}^{\left(\alpha\right),n+1}\rho_{i}^{\left(\alpha\right),n+1}\right)^{-1}+\frac{1}{2}\left(u_{i}^{n+1}\right)^{2},\label{eq:conservative energy update-1}
\end{multline}
 where the pressure $\tilde{p}_{i}^{\left(\alpha\right)}$ is calculated in a special way to increase robustness and sometimes aid in maintaining positivity \cite{Cutforth:2021} via
\[
\tilde{p}_{i}^{\left(\alpha\right)}\leftarrow\begin{cases}
p_{i}^{\left(\alpha\right),n+1}, & u_{i+1/2}^{n+1/2}-u_{i-1/2}^{n+1/2}<0\text{ i.~e. }\left(\boldsymbol{\nabla}\cdot\boldsymbol{u}\right)<0\text{ (compression with growing }p\text{)}\\
\overline{p}_{i}^{\left(\alpha\right),n}, & u_{i+1/2}^{n+1/2}-u_{i-1/2}^{n+1/2}\geqslant0\text{ i.~e. }\left(\boldsymbol{\nabla}\cdot\boldsymbol{u}\right)\geqslant0\text{ (expansion with decreasing }p\text{)},
\end{cases}
\]
 and $p_{i}^{\left(\alpha\right),n+1}$ is, in turn, found by minimizing $F\left[e_{i}^{\left(\alpha\right),n+1}\right]=$
\begin{multline}
f_{i}^{\left(\alpha\right),n+1}\rho_{i}^{\left(\alpha\right),n+1}e_{i}^{\left(\alpha\right),n+1}
-f_{i}^{\left(\alpha\right),n}\rho_{i}^{\left(\alpha\right),n}e_{i}^{\left(\alpha\right),n}
+P^{\left(\alpha\right)}\left(\rho_{i}^{\left(\alpha\right),n+1},e_{i}^{\left(\alpha\right),n+1}\right)\cdot\left(f_{i}^{\left(\alpha\right),n+1}-\tilde{f}_{i}^{\left(\alpha\right),n+1}\right)\\
+\left(V_{i+1/2}^{\left(\alpha\right),\text{advected}}\tilde{\rho}_{i+1/2}^{\left(\alpha\right),n+1/2}\tilde{e}_{i+1/2}^{\left(\alpha\right),n+1/2}-V_{i-1/2}^{\left(\alpha\right),\text{advected}}\tilde{\rho}_{i-1/2}^{\left(\alpha\right),n+1/2}\tilde{e}_{i-1/2}^{\left(\alpha\right),n+1/2}\right)\rightarrow0.
\end{multline}
 The pressure $P^{\left(\alpha\right)}\left(\rho_{i}^{\left(\alpha\right),n+1},e_{i}^{\left(\alpha\right),n+1}\right)$ is calculated from the corresponding equation of state.
Compute the total energy in all other cells directly --- as follows:
\begin{multline}
E_{i}^{\left(\alpha\right),n+1}\leftarrow\left(f_{i}^{\left(\alpha\right),n+1}\rho_{i}^{\left(\alpha\right),n+1}\right)^{-1}f_{i}^{\left(\alpha\right),n}\overline{\rho}_{i}^{n}E_{i}^{\left(\alpha\right),n}\\
-\Biggl(\left(f_{i}^{\left(\alpha\right),n+1}\rho_{i}^{\left(\alpha\right),n+1}\right)^{-1}V_{i+1/2}^{\left(\alpha\right),\text{advected}}\tilde{\rho}_{i+1/2}^{\left(\alpha\right),n+1/2}\tilde{E}_{i+1/2}^{\left(\alpha\right),n+1/2}\\
-\left(f_{i}^{\left(\alpha\right),n+1}\rho_{i}^{\left(\alpha\right),n+1}\right)^{-1}V_{i-1/2}^{\left(\alpha\right),\text{advected}}\tilde{\rho}_{i-1/2}^{\left(\alpha\right),n+1/2}\tilde{E}_{i-1/2}^{\left(\alpha\right),n+1/2}\Biggr)\\
-\frac{\Delta t}{\Delta x}u_{i}^{n+1/2}\left(\frac{f_{i}^{\left(\alpha\right),n+1}\rho_{i}^{\left(\alpha\right),n+1}}{\overline{\rho}_{i}^{n+1}}\left(\bar{p}_{i+1/2}^{n+1/2}-\bar{p}_{i-1/2}^{n+1/2}\right)\right)\left(f_{i}^{\left(\alpha\right),n+1}\rho_{i}^{\left(\alpha\right),n+1}\right)^{-1}\\
-\overline{p}_{i}^{n+1/2}\underbrace{\left(\left(1-\sum_{\beta=1}^{N_{f}}\tilde{f}_{i}^{\left(\beta\right),n+1}\right)\frac{\tilde{f}_{i}^{\left(\alpha\right),n+1}\left(\sum_{\beta=1}^{N_{f}}\tilde{f}_{i}^{\left(\beta\right),n+1}/\left(\tilde{K}_{S}^{\left(\beta\right)}\right)_{i}^{n+1}\right)^{-1}}{\left(\tilde{K}_{S}^{\left(\alpha\right)}\right)_{i}^{n+1}}\right)}_{\substack{\shortparallel\\
f_{i}^{\left(\alpha\right),n+1}-\tilde{f}_{i}^{\left(\alpha\right),n+1}
}
}\left(f_{i}^{\left(\alpha\right),n+1}\rho_{i}^{\left(\alpha\right),n+1}\right)^{-1}.\label{eq:conservative energy update}
\end{multline}
\item Update the pressures and bulk moduli of the phases using their equations of state:
\[
p_{i}^{\left(\alpha\right),n+1}\stackrel{\text{EoS }\left(\alpha\right)}{\leftarrow}P^{\left(\alpha\right)}\left(\rho_{i}^{\left(\alpha\right),n+1},E_{i}^{\left(\alpha\right),n+1}-\frac{1}{2}u_{i}^{n+1}\right),
\]
\[
\left(K_{S}^{\left(\alpha\right)}\right)_{i}^{n+1}\stackrel{\text{EoS }\left(\alpha\right)}{\leftarrow}K_{S}^{\left(\alpha\right)}\left(\rho_{i}^{\left(\alpha\right),n+1},E_{i}^{\left(\alpha\right),n+1}-\frac{1}{2}u_{i}^{n+1}\right).
\]
\item Perform an isentropic relaxation towards a common effective pressure $\bar{p}_{i}^{n+1}$ in mixed cells under the constraint of $\sum_{\alpha=1}^{N_{f}}f_{i}^{\left(\alpha\right)}=1$. Since the multiple phases in mixed cells can lose mechanical equilibrium after update steps 8--11, volume fractions, densities, and energies are iteratively redistributed between phases within a cell until equilibrium is reached again. Redistribution is determined from the following equations
for linearised changes $ \Delta f^{\left(\alpha\right),n+1} $ in these fractions according to the procedure introduced in the work \cite{Law:2024}:
\[
\bar{p}_{i}^{n+1}=p_{i}^{\left(\alpha\right),n+1}+\Delta p_{i}^{\left(\alpha\right),n+1}=p_{i}^{\left(\alpha\right),n+1}-\frac{K_{S}^{\left(\alpha\right),n+1}}{f_{i}^{\left(\alpha\right),n+1}}\Delta f_{i}^{\left(\alpha\right),n+1},
\]
\[
\sum_{\alpha=1}^{N_{f}}\Delta f_{i}^{\left(\alpha\right),n+1}=0.
\]

The parameter limiting relative compression is set to $ \delta_{-} = 0.01 $, to ensure volume fraction changes during relaxation remain sufficiently small for the linearisation. The convergence condition for the relaxation iterations is implemented based on the error in
the $L_{1}$ norm:
\[
\sum_{\left(\alpha\right)}\left|\bar{p}_{i}^{n+1}-p_{i}^{\left(\alpha\right),n+1}\right|<N_{fi}\max\left\{ 10^{-4},10^{-4}\left|\bar{p}_{i}^{n+1}\right|\right\},
\]
 where $N_{fi}$ stands for the quantity of phases present in the $i$th cell at the time of the current relaxation iteration.
\end{enumerate}

That is the general algorithm. Concerning steps 5 and 3, we recommend the use of the HLLC ({\textit{Harten--Lax--van Leer--Contact}}) \cite{Toro:1994}  
solver and propose to pair it either with the values obtained from the  MUSCL-Hancock-type reconstruction ({\textit{Monotonic
Upstream-centred Scheme for Conservation Laws}} with temporal reconstruction
by Steve Hancock) \cite{vanLeer:1979,Colella:1982} used in \cite{Miller:1996}
or with the left upwind ($\bar{\rho}_{i-1/2}^{\ell}=\bar{\rho}_{i-1}$
etc.) and right upwind ($\bar{\rho}_{i-1/2}^{r}=\bar{\rho}_{i}$
etc.) values.

Second-order MUSCL-Hancock reconstruction consists of
limited local extrapolation with linear Taylor expansions to $t=t^{n}+\Delta t\,/\,2$,
$x=x_{i}\pm\Delta x\,/\,2$ along the filtered upwind characteristics from the
averages at the cell centres \cite{vanLeer:1979,Miller:1996}.

\enlargethispage{1\baselineskip}
Velocities and pressures in step 5 of the algorithm are then determined from the reconstructed values or from the upwind values in the following way \cite{Law:2024}:
\[
u_{i+1/2}^{n+1/2}=\begin{cases}
u_{i+1/2}^{\ell\,n}, & S_{L}\geqslant0\\
S_{\ast}, & S_{L}<0<S_{R}\\
u_{i+1/2}^{r\,n}, & S_{R}\leqslant0,
\end{cases}
\]
\[
\bar{p}_{i+1/2}^{n+1/2}=\begin{cases}
\bar{p}_{i+1/2}^{\ell\,n}, & 0\leqslant S_{L}\\
\bar{p}_{i+1/2}^{\ell\,n}+\bar{\rho}_{i+1/2}^{\ell\,n}\left(S_{L}-u_{i+1/2}^{\ell\,n}\right)\left(S_{\ast}-u_{i+1/2}^{\ell\,n}\right), & S_{L}<0\leqslant S_{\ast}\\
\bar{p}_{i+1/2}^{r\,n}+\bar{\rho}_{i+1/2}^{r\,n}\left(S_{R}-u_{i+1/2}^{r\,n}\right)\left(S_{\ast}-u_{i+1/2}^{r\,n}\right), & S_{\ast}<0\leqslant S_{R}\\
\bar{p}_{i+1/2}^{r\,n}, & S_{R}<0,
\end{cases}
\]
 where $S_L$, $S_R$ are the wavespeeds, computed as in \cite{Cutforth:2021},
 and $ S_{\ast} $ is the HLLC speed of the contact discontinuity, computed in the standard way \cite{Toro:1994}.

An alternative method of resolving the wave formation between neighbouring cells (to finding $ u^{n+1/2}_{i+1/2} $, $ \bar{p}^{n+1/2}_{i+1/2} $) is described in \cite{Menshov:2022en}, with the method supporting virtually non-diffusive contact-interface capturing. The method is based on the solution of non-self-similar composite Riemann problems explicitly including the phase boundaries.

\section{Results}

We shall now describe the process of wave formation effected by the collision of the metal plates.
At the initial instant $ t=0 $, the coordinate of their contact interface is $ x = \unit[5]{mm} $. The collision of the lead plate with the steel plate at that initial instant triggers the formation of two shock waves propagating from the contact discontinuity into both of the metals.
The shock waves formed in the metals move until they reach the free surfaces of the plates --- that is, their phase boundaries with the surrounding air. Let us first consider the process of wave propagation in the steel plate. At the time $ t \sim \unit[0.58]{\mus} $,
the shock wave moving to the left within the steel plate reaches the edge of the plate. Its interaction with the air creates a right-moving unloading wave. A similar process occurs in the lead plate; the unloading wave at the free surface is formed at time $ t \sim \unit[0.89]{\mus} $
and propagates to the left. At time $ t \sim \unit[1.13]{\mus} $,
the unloading wave in steel reaches the contact interface between the plates and passes through it, thereby increasing the absolute value of the contact velocity. By the time $ t \sim \unit[1.3]{\mus} $, the metal contact speed has increased from $ \unit[191]{m/s} $ to $ \unit[425.2]{m/s} $.
At $ t \sim \unit[1.4]{\mus} $, the unloading waves intersect inside the lead plate, interact with each other, form negative pressures and cause a decrease in density at the point of intersection. By the time $ t \sim \unit[1.72]{\mus} $,
the unloading wave formed at the right edge of the lead plate has reached the interface with the steel plate, causing the speed of the interface to start decreasing.

\begin{figure}[H]
\begin{center}
{\large\textbf{
  \includegraphics[scale=1]{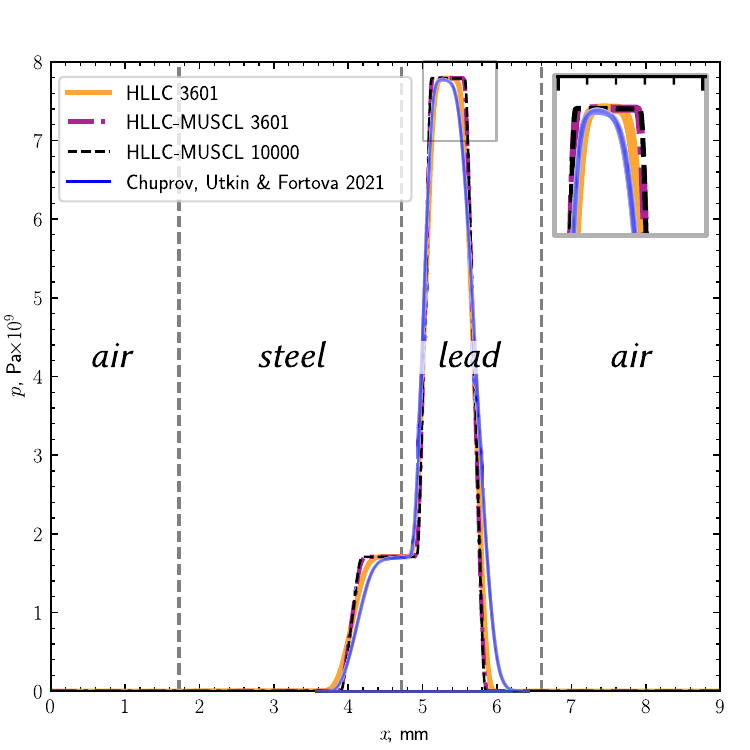}
}}{\large\par}
\par\end{center}
\starcaption{
Pressure profiles at time $ t=\unit[1.3]{\mus} $, obtained by various methods. The thin solid line corresponds to data from \protect\cite{Chuprov:2021}. The dashed black line corresponds to data calculated by HLLC with MUSCL reconstruction on a grid of $10\,000$ cells. The thick solid and pale dashed lines correspond to data calculated by the current method with and without reconstruction, with the same spatial resolution as the method in \protect\cite{Chuprov:2021}. The interfaces between phases are indicated by dotted grey vertical lines. The inset in the upper right corner shows an enlarged plot of the pressure field on the $x$-axis segment with $ x \in \left[ \unit[5]{mm} \mathop{..} \unit[6]{mm} \right] $
}\label{fig:3}
\end{figure}

The wave-formation process described above is reproduced in this work by direct numerical simulation using a finite-volume scheme with the VOF method, both with and without MUSCL-Hancock reconstruction. A comparison of pressure profiles at time $ t \sim \unit[1.3]{\mus} $ obtained by different methods is shown in Fig.\ \ref{fig:3}.
By this time, the unloading wave from steel has finished interacting with the interface of the plates. The short-dashed black line in the figure shows the numerical solution for pressure obtained on a detailed grid of $ 10\,000 $ cells using MUSCL-Hancock reconstruction with $ \text{CFL} = 0.2 $. The thicker long-dashed and thick solid lines correspond to the pressure profiles calculated with and without MUSCL-Hancock reconstruction, respectively, on a grid of $ 3\,601 $ cells.
A comparison of the plots leads to the conclusion that the numerical approach with reconstruction has higher resolution. Pressure profiles without reconstruction are subject to noticeable numerical diffusion compared to those with reconstruction.
Similar results on a grid of $ 3\,601 $ cells
were obtained in \cite{Chuprov:2021} using the Baer--Nunziato model. The corresponding pressure profile is shown as a thin solid line in Fig.\ \ref{fig:3}.

\begin{figure}[H]
\begin{center}
{\large\textbf{
  \includegraphics[scale=0.9]{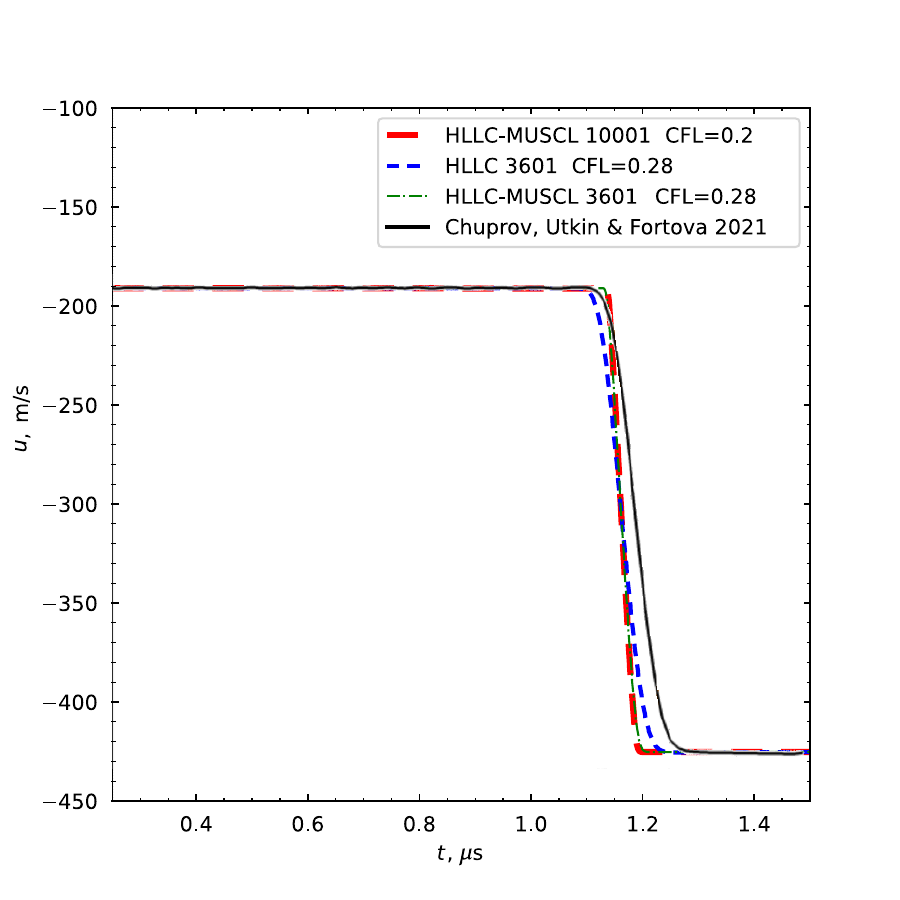}
}}{\large\par}
\par\end{center}
\starcaption{
Plot of the metallic-plate interface velocity versus time, obtained by various methods. The solid black line corresponds to the data from \protect\cite{Chuprov:2021}.
The thickest line --- made up by sparsely spaced dashes --- corresponds to data obtained with HLLC and MUSCL reconstruction on a grid of $10\,001$ cells. The less thick dashed
and thin dash-dot lines correspond to data obtained with the current method without
and with reconstruction with the same spatial resolution as the method
in \protect\cite{Chuprov:2021}
}\label{fig:4}
\end{figure}

Fig.\ \ref{fig:4}
shows plots of the dependence of the contact-interface velocity for metal plates on time in the first $ \unit[1.5]{\mus} $. The interface velocity in our numerical solution is taken as the velocity in the mixed
steel/lead cell. The thickest dashed line corresponds to the calculation with $ 10\,000 $ cells and MUSCL-Hancock reconstruction with $ \text{CFL} = 0.2 $.  The thin dash-dot and slightly thicker dashed lines show the velocity plots calculated on a grid of $ 3\,601 $ cells and at $ \text{CFL} = 0.28 $ with and without reconstruction. The thin solid black line is a plot from \cite{Chuprov:2021}. Due to the low diffusion of the numerical scheme, the interface-velocity plot obtained with reconstruction on a grid of $ 3\,601 $ cells already agrees to within $ 4.6 \% $ with the velocity plot on a detailed grid. At the same time, the calculation using the method described above shows a faster settling time (stabilisation) for the interface velocity ($ t \sim \unit[1.2]{\mus} $) compared to the work \cite{Chuprov:2021}.
Note that the time of arrival of the unloading wave from steel to the contact interface when calculated using the described method is approximately $\unit[1.13]{\mus}$. This is consistent with the experimental data \cite{Yakovlev:1975}
and the numerically obtained value of $\unit[1.05]{\mus}$ from \cite{Chuprov:2021}.

\begin{figure}[H]
\begin{center}
{\large\textbf{
  \includegraphics[scale=0.80]{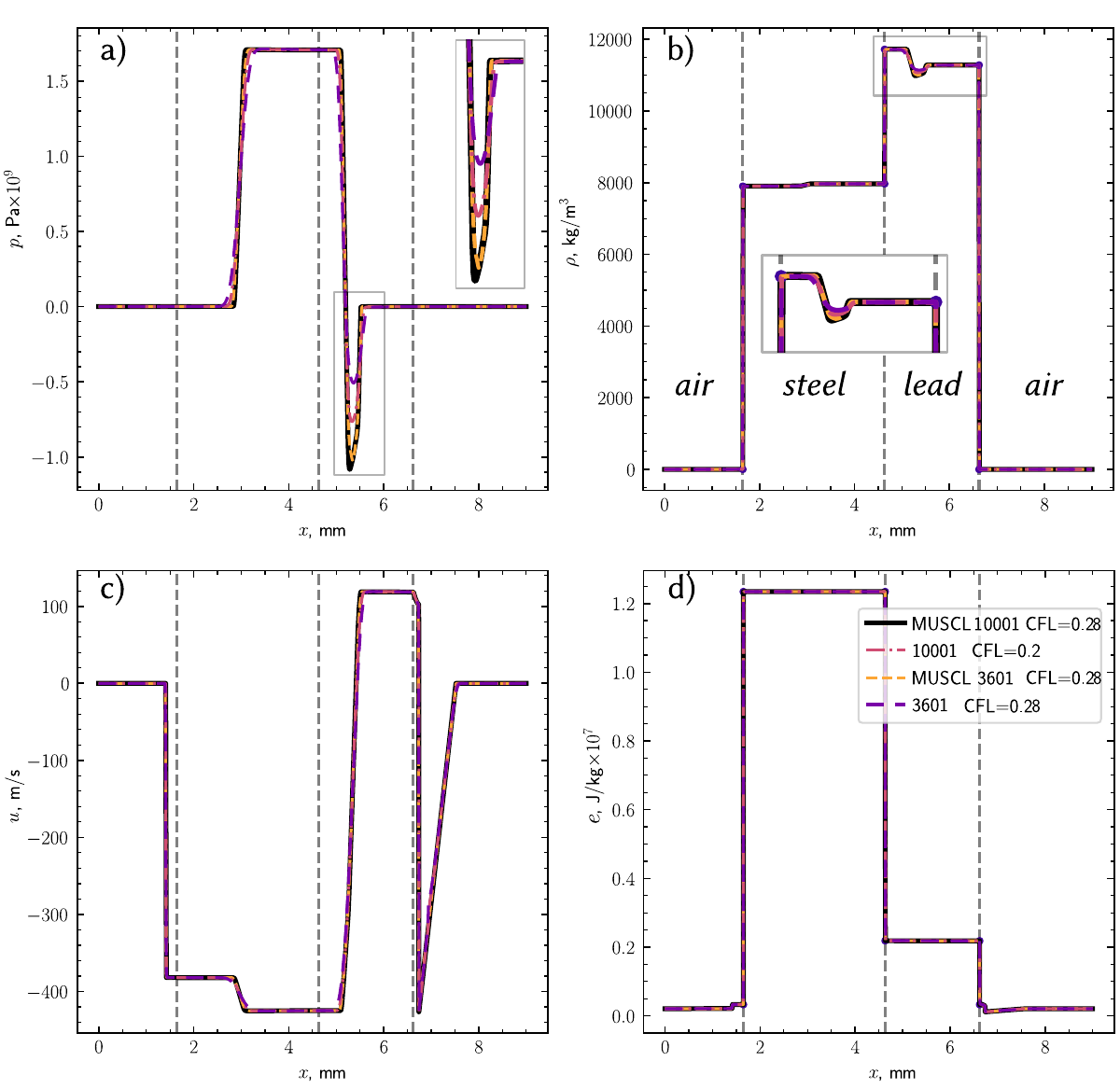}
}}{\large\par}
\par\end{center}
\starcaption{
Profiles for pressure $ p $ (a), density $ \rho $ (b), velocity $ u $ (c) and internal specific energy $ e $ (d) at time $t=\unit[1.5]{\mus}$, obtained with the current method, with and without reconstruction.
The plots calculated using MUSCL at $10\,001$ cells with CFL$=0.28$ are shown by a thick solid black line. The palest densely dashed line represents the solution obtained using MUSCL at $3\,601$
cells; the dashed line with larger dash spacing is the solution obtained without
MUSCL at the same resolution, and the dash-dot line is the solution without MUSCL at $10\,001$ cells with CFL$=0.2$. The
 bold dashed grey vertical lines correspond to the phase interfaces
}\label{fig:5}
\end{figure}

\enlargethispage{1\baselineskip}
After $ t\sim\unit[1.5]{\mus} $,
negative pressure occurs in the lead phase. This fact impairs further high-quality hydrodynamic modelling \cite{Chuprov:2021}. Our proposed numerical method is able to overcome the associated difficulties: the profiles of pressure, density, velocity and specific internal energy at time $ t\sim\unit[1.5]{\mus} $ obtained with the method are shown in Fig.\ \ref{fig:5}a)--d).
The superiority of the reconstruction-based method in a region of pressure drop is displayed in the inset of the pressure plot (Fig.\ \ref{fig:5}a)). A similar situation is observed with the dip in the density plots (Fig.\ \ref{fig:5}b)).

When formulae \eqref{eq:conservative energy update-1} for internal energies are used in cells that are mixed or adjacent to mixed cells, this introduces conservation error into the numerical scheme. The peak of the error occurs after $ t\approx\unit[0.8965]{\mus} $. The total error up to $ t = \unit[2]{\mus} $ is of the order of $0.06\%$ for the effective density $\sum_{i=0}^{N_{x}-1}\bar{\rho}_{i}^{n}$ and $0.02\%$ for the integral total effective energy $\sum_{i=0}^{N_{x}-1}\left(\bar{\rho}\bar{E}\right)_{i}^{n}$ when calculated on a grid of
$3\,601$ cells with $\text{CFL}=0.28$. 

\section{Conclusions}

Multifluid hydrodynamic modelling with interface tracking by the VOF method is applied to the investigation of a one-dimensional problem of high-speed collision between steel and lead plates. Owing to the use of the Eulerian approach with the HLLC solver, the model has low numerical diffusion and allows contact interfaces of immiscible phases to be resolved with high accuracy. The numerical algorithm seamlessly resolves tensile stresses without requiring additional procedures to account for negative pressures. In the future, generalising the method to the two-dimensional formulation will allow us to continue studying the instability of the contact interface between colliding plates in explosion welding.

\section*{Acknowledgements}

The authors are grateful to Academician Mikhail Aleksandrovich Guzev, %
Dr\ M.~E.\ Povarnitsyn, and Dr\ P.~A.\ Chuprov for useful discussions.

We also wish to acknowledge Dr\ Vadim Vladimirovich Shepelev for inspiring this work and contributing to its foundational ideas. His insights were invaluable in shaping the research direction.

\section*{Funding}

The present work was supported by the Ministry of Science and Higher Education of the Russian Federation within the framework of the Russian State Assignment of the ICAD RAS No.\ 124022400174-3.

\section*{Conflict of Interest}
The authors declare that they have no conflicts of interest.

\begin{center}
\begin{singlespace}
\sloppy
\addcontentsline{toc}{section}{References}
\printbibliography
\fussy
\end{singlespace}
\end{center}
\end{document}
%
%
%